# Anomalous effects in conductivity of thin film samples obtained by thermolysis of PVC in solution

V. I. Kryshtob, L. A. Apresyan, D. V. Vlasov, and T.D.Vlasova

Prokhorov General Physics Institute, Russian Academy of Science

ul.Vavilova 38, Moscow, 119991 Russia

**ABSTRACT**

It was stated earlier that in the samples of PVC films plasticized by a new type of plasticizer (modifier-A) abnormal transitions are possible from the state of low electrical conductivity to the state of high conductivity. For the first time a similar anomalous transition is demonstrated in thin-film samples of (partially dehydrochlorinated) PVC obtained by thermolysis in solution, without use of any special additives, plasticizers, ionizing radiation, etc. Anomalous effects of conductivity in the experimental data obtained clearly correlate with a concentration of conjugated double bonds (polyvinylen fragments) in the main chain of thermalized PVC macromolecules.

ВВЕДЕНИЕ.

Известно, что полимеры, содержащие систему сопряженных двойных связей, как правило, относятся к классу электропроводящих систем. При этом специфика их электронного строения такова, что способна придавать им качественно новые свойства, такие как «полупроводниковая» проводимость, фотоэлектрическая чувствительность, наличие парамагнитных центров и т.д. [1]
Использование процесса термолиза ПВХ в растворе, позволяет осуществить плавный переход от стабильного изолятора ПВХ к полимеру с двойными сопряженными связями, обладающим «полупроводниковым» уровнем проводимости. Если для ПВХ аномалии проводимости достаточно подробно исследованы, то для промежуточных случаев данных об аномальных переключениях электропроводности практически отсутствуют.

В частности, известно, что в полиацетилене (содержащем 100% двойных связей и являющимся прототипом электропроводящих полимеров) возможно проявление аномально высокой (вплоть до металлической) электропроводности, причем подобные переходы наблюдаются лишь в случае допирования полимера электронно-донорными или электронно-акцепторными соединениями [1].

Ранее авторами было показано [2,3], что при использовании вместо традиционных типов пластификаторов модификаторов нового типа (модификаторов типа А) при изменении массовой доли последнего в ПВХ наряду с традиционным (монотонным) и незначительным увеличением электропроводности наблюдаются аномальные переходы в состояние высокой проводимости (СВП). При этом изменение электропроводности составляет 3-4 порядков и более, уменьшая оценку удельной электропроводности до $10^3$ Ом·см. Скачкообразные переключения электропроводности переводят модифицированный ПВХ из разряда типичных диэлектриков в разряд веществ с уровнем проводимости характерным для полупроводников [2-8].

В данной работе исходные образцы ПВХ подвергались химической модификации методом термолиза последнего в растворе, без использования при этом каких-либо пластификаторов, специальных добавок, ионизирующих излучений и т.д., причем аномалии электропроводности можно было наблюдать для различных стадий развития процесса дегидрохлорирования образцов в растворе.

## ЭКСПЕРИМЕНТАЛЬНАЯ ЧАСТЬ

Опытные образцы частично дегидрохлорированного ПВХ методом термолиза в растворе получали следующим образом.
Вначале получали 4% раствор ПВХ (марки С-70) в растворителе (ацетофеноне). Растворение ПВХ осуществляли при перемешивании при комнатной температуре в течение 12 часов до получения гомогенного прозрачного раствора. В дальнейшем раствор помещался в пробирку и подвергался термолизу при Т=190$^0$С в течение 20-480минут. Порция , необходимая для получения образца пленки заливалась на стеклянную подложку и подвергалась сушке при Т=95$^0$С в термошкафу в течение 48 часов. Далее полученные образцы подвергались визуальному, технологическому и органолептическому контролю (прозрачность, цвет, легкость снятия с подложки, прочность, изгибостойкость, залипаемость и т.д.).

Измерения образцов ПВХ-пленок по показаниям электропроводности осуществляли на приборе, описанном [2-8] с ГОСТированной кольцевой измерительной ячейкой.

## РЕЗУЛЬТАТЫ И ИХ ОБСУЖДЕНИЕ

Данные по основным электрофизическим свойствам (условно рассчитанным на удельное объемное электрическое сопротивление $\rho_v$ , Ом см) полученных в лабораторных условиях образцов сополимеров винилена и винилхлорида (частично дегидрохлорированных образцов исходного ПВХ) представлены в Таблице.

ТАБЛИЦА

| (1) | (2) | (3) | (4) | (5) | (6) | (7) |
|---|---|---|---|---|---|---|
| № образца | Время дегидрохлорирования (термолиза) ПВХ в растворе, мин. | Толщина пленок, мкм | Внешний вид и органолептические свойства | Показатель уд.об. сопр., $\rho_v$ ,Ом см (в СНП) | Показатель уд.об. сопр., $\rho_v$ Ом см (в СВП) | Примечание |
| №1 (исх. ПВХ) | 0 | 15 | Бесцветная, прозрачная, хрупкая, ломающаяся пленка | $10^{15}$ | Отсутвт. | Из данных [9],с.809. |
| №2 | 20 | 11 | Прозр., менее хрупкая пленка слабо-желтого цвета | $2,8 \cdot 10^{14}$ | Отсутств. | Переход в СВП не набл. |
| №3 | 240 | 10 | Прозр., желтого цвета, прочная, незалипающая, гнущаяся, легко снимающаяся со стекл.подл. пленка | $7.1 \cdot 10^{13}$ | $<8,2 \times 10^3$ | Наблюдается переход в СВП |
| №4 | 320 | 10 | -«- | $4,5 \cdot 10^{13}$ | $<8,6 \times 10^3$ | -«- |

| №5 | 480 | 12 | Прозр., более насыщенного желтого цвета, незалипающая, легко снимающаяся со стекл.подл., гнущаяся пленка | $4,0 \cdot 10^{13}$ | $<8,0 \times 10^3$ | Наиболее легко переходящая в СВП |

Как в случае образцов ПВХ, модифицированных новым типом пластификатора (модификатором А) [8], часть полученных методом термолиза из раствора образцов сополимера винилена и винилхлорида обладала впервые наблюдаемым аномальным свойством: переходом их состояния нормальной проводимости (СНП) в состояние высокой проводимости (СВП) без использования при этом дополнительной операции допирования.

При этом, в отличие от образцов ПВХ, пластифицированных модификатором типа А [3-8] разница максимальных и минимальных значений удельного объемного сопротивления ($\rho_v$) для частично дегидрохлорированных образцов ПВХ при переходах из СНП в СВП составляла уже более 10 порядков и в абсолютном значении составляло в СВП величину $\rho_v < 8 \cdot 10^3$ Ом см.

Отметим также, что переход в СВП частично дегидрохлорированных образцов ПВХ осуществлялся гораздо легче в случае использования большего времени термолиза (а значит предположительно и большей степени дегидрохлорирования исходных образцов ПВХ).

Сопоставительный визуальный, органолептический и технологический анализ полученных образцов позволяет дополнительно сделать следующие предварительные выводы:

- степень дегидрохлорирования образцов невысока ( н/б 10-15%); при этом число последовательных сопряженных связей в блоке поливинилена не превышает 3-4 [10];

- схожесть органолептических и технологических свойств полученных образцов с образцами [8] позволяет допустить, что полиеновые фрагменты в образцах непластифицированного исходного ПВХ наряду с аномальным проявлением электропроводящих свойств могут играть дополнительно роль «пластификатора», т.е. способствовать в некоторой степени улучшению комплекса его физико-механических и технологических свойств.

С учетом вышеизложенного, а также того обстоятельства, что предположительно механизмы электропроводности образцов ПВХ, пластифицированных модификатором А и полученных путем термолиза в растворе (т.е. частично дегидрохлироллванных) должны существенно отличаться, использование термолизованных образцов ПВХ оказывается более предпочтительным как минимум по причине имеющейся принципиальной возможности улучшения электропроводящих свойств образцов, содержащих полиеновые структуры за счет дополнительного использования операций допирования

.

## ЛИТЕРАТУРА